\documentclass[conference]{IEEEtran}
\IEEEoverridecommandlockouts
\usepackage{cite}
\usepackage{url}
\usepackage{stfloats}
\usepackage[font=footnotesize]{caption}
\usepackage{subcaption}
\usepackage{amsmath,amssymb,amsfonts}
\usepackage{bm}
\usepackage{algorithm}
\usepackage{algpseudocode}
\algrenewcommand{\algorithmiccomment}[1]{\textit{{\footnotesize \% #1}}}
\usepackage{graphicx}
\usepackage{textcomp}
\usepackage{xcolor}
\usepackage[utf8]{inputenc}
\usepackage[normalem]{ulem}
\usepackage{soul}
\usepackage{ulem}
\def\BibTeX{{\rm B\kern-.05em{\sc i\kern-.025em b}\kern-.08em
    T\kern-.1667em\lower.7ex\hbox{E}\kern-.125emX}}
\usepackage{acronym}
\usepackage{multicol}
\usepackage{multirow}
\usepackage{hyperref}
\usepackage[capitalise]{cleveref}
\usepackage[letterpaper, left=0.625in, right=0.625in, bottom=1in, top=0.75in]{geometry}

\acrodef{rsm}[RSM]{receive spatial modulation}
\acrodef{adc}[ADCs]{analog-to-digital converters}
\acrodef{zf}[ZF]{zero-forzing}
\acrodef{ras}[RAS]{receiver antenna selection}
\acrodef{ara}[ARA]{active receiver antenna}
\acrodef{aras}[ARAs]{active receiver antennas}
\acrodef{dl}[DL]{downlink}
\acrodef{ho}[HO]{Handover}
\acrodef{rsm}[RSM]{receive spatial modulation}
\acrodef{adc}[ADC]{analog-to-digital converter}
\acrodef{riss}[RISs]{Reconfigurable intelligent surfaces}
\acrodef{ris}[RIS]{reconfigurable intelligent surface}
\acrodef{snr}[SNR]{signal-to-noise ratio}
\acrodef{sinr}[SINR]{signal-to-interference-plus-noise ratio}

\acrodef{rf}[RF]{radio-frequency}
\acrodef{los}[LOS]{line-of-sight}
\acrodef{nlos}[NLOS]{non-line-of-sight}
\acrodef{nf}[NF]{near-field}
\acrodef{ff}[FF]{far-field}
\acrodef{spd}[SPD]{spatial power density}
\acrodef{ml}[ML] {maximum-likelihood}
\acrodef{lis}[LIS] {large intelligent surface}
\acrodef{mimo}[MIMO] {multiple-input-multiple-output}
\acrodef{simo}[SIMO]{single-input-multiple-output}
\acrodef{mvdr}[MVDR] {minimum variance distortionless response}
\acrodef{bs}[BS] {base station}
\acrodef{cmab}[CMAB] {contextual multi-armed bandit}
\acrodef{cmabs}[CMABs] {Contextual multi-armed bandits}
\acrodef{mab}[MAB] {multi-armed bandit}
\acrodef{lte}[LTE]{long-term evolution}
\acrodef{lmmse}[LMMSE]{linear minimum mean square error}
\acrodef{nr}[NR]{New Radio}
\acrodef{cho}[CHO]{conditional handover}
\acrodef{daps}[DAPS]{dual active protocol stack}
\acrodef{ltm}[LTM]{low-layer triggered mobility}
\acrodef{ppo}[PPO]{proximal policy optimization}
\acrodef{dqn}[DQN]{deep Q-networks}
\acrodef{ssb}[SSB] {synchronization signal block}
\acrodef{ssbs}[SSBs] {synchronization signal blocks}
\acrodef{mcs}[MCS] {modulation and coding scheme}
\acrodef{pp}[PP] {ping-pont}
\acrodef{rsrp}[RSRP]{reference signal received power}
\acrodef{hof}[HOF]{HO failures}
\acrodef{rlf}[RLF]{Radio link failures}
\acrodef{rss}[RSS]{received signal strength}

\DeclareMathOperator*{\argmin}{arg\,min}

\DeclareUnicodeCharacter{2061}{}

\newcommand{\Hbr}{\mathbf{H}_{br}}

\title{Contextual Bandits and Reconfigurable Intelligent Surfaces for Predictive LTM Handover Decisions\\
\thanks{This work was funded through the project 5GSmartFact, funded by the European Union's H2020 research and innovation program
under the MSCA grant ID 956670, project 6-SENSES grant PID2022-138648OB-I00, funded by MCIN/AEI/10.13039/501100011033 and by FEDER-UE, ERDF-EU \textit{A way of making Europe}, and the grant 22CO1/008248.}} 

\author{\IEEEauthorblockN{Ainna Yue Moreno-Locubiche, Josep Vidal, Olga Muñoz-Medina, Margarita Cabrera-Bean}
\IEEEauthorblockA{\textit{Dept. of Signal Theory and Communications, Universitat Politècnica de Catalunya - BarcelonaTech, Spain} \\
\{ainna.yue.moreno, josep.vidal, olga.muñoz, marga.cabrera\}@upc.edu}}

\begin{document}
\maketitle
\begin{abstract}
This article addresses the challenge of optimizing handover (HO) in next-generation wireless networks by integrating Reconfigurable Intelligent Surfaces (RIS), predicting received signal power, and utilizing learning-based decision-making. A conventional reactive HO mechanism, such as lower-layer triggered mobility (LTM), is enhanced through linear prediction to anticipate link degradation. Additionally, the use of RIS helps to mitigate signal blockage and extend coverage. An online trained non-linear Contextual Multi-Armed Bandit (CMAB) agent selects target gNBs based on context features, which reduces unnecessary HO and signaling overhead. Extensive simulations evaluate eight combinations of these techniques under realistic mobility and channel conditions. Results show that CMAB and RSRP prediction consistently reduce the number of HO, ping-pong rate and cell preparations, while RIS improves link reliability.
\end{abstract}


\section{Introduction} \label{Intro}
\ac{ho} mechanisms in mobile networks are designed to maintain seamless connectivity as users move between cells. The rise of 6G networks, with its higher frequency bands and more stringent service quality standards, introduces new challenges in maintaining seamless connectivity, especially in dense urban areas with \ac{nlos} conditions, and in high-mobility environments such as urban transport, UAVs, and vehicular networks \cite{giordani2020toward}. Traditional \ac{ho} mechanisms, which rely on reactive signal strength measurements, often fail in these dynamic scenarios, leading to frequent interruptions, high signaling overhead, and degraded service quality \cite{twc2024.3471393}. These limitations reduce service reliability and increase latency. To address these limitations, we propose a novel framework that integrates three cutting-edge technologies:

\begin{itemize}
    \item \textbf{\ac{cmabs}:} A reinforcement learning-based decision engine that adapts to real-time network conditions, learning best \ac{ho} decision policies from experience.
    \item \textbf{Prediction of received signal power:} A \ac{lmmse} predictor that anticipates future received signal power, enabling proactive \ac{ho} decisions and reducing \ac{pp} effects.
    \item \textbf{\ac{riss}:} Passive programmable metasurfaces that reflect and redirect signals to improve coverage and reliability in \ac{nlos} conditions.
\end{itemize}

While the usage of \ac{riss} has demonstrated potential in enhancing spectral and energy efficiency, its integration into mobility management remains underexplored. This work bridges that gap by combining the usage of \ac{riss} with predictive and learning-based algorithms to improve handover robustness in dense, fast-changing environments. The proposed system aligns with recent 3GPP advancements but goes further by embedding learning into the \ac{ho} process, transforming it from reactive to anticipatory. The result is a significant improvement in HO performance indicators.

\subsection{Literature review}
Conventional \ac{ho} procedures, which rely on signal strength thresholds, often fail in dense urban environments and high-mobility scenarios.
In ultra-dense networks, overlapping cell coverage leads to frequent handovers, causing \ac{pp} effects, unnecessary resource reservation, increased signaling overhead and radio link failures.
These procedures include \cite{giordani2020toward}: hard \ac{ho} (break-before-make) used in \ac{lte} and \ac{nr}, and soft \ac{ho} (make-before-break) found in WCDMA, with hybrid approaches combining aspects of both. More advanced mechanisms in 5G and beyond include \ac{cho}, where the UE executes transitions based on pre-set conditions; \ac{daps}, which allows simultaneous connections to source and target cells during \ac{ho}; and \ac{ltm}, which enables ultra-fast switching using MAC and PHY layers \cite{ericsson_l1_l2_triggered_mobility}. These methods aim to reduce latency and improve reliability but remain non-adaptive rule-based \cite{survey_handover_5g_nr,3gpp_38300}.

By leveraging AI-driven models, beamforming optimization, and \ac{ris}-assisted strategies, researchers aim to address the challenges posed by harsh mobility situations, dense deployments, and dynamic channel conditions.
Machine learning offers promising solutions for dynamic HO management and outperforms static approaches. Deep learning models like LSTM can predict user movement \cite{alsabah2021handovers}, but are trained with large historical data sets for fixed terminal speeds. Reinforcement learning algorithms such as \ac{dqn}  and \ac{ppo}  adapt \ac{ho}  parameters in real time \cite{liu2021reinforcement,voigt2025deepreinforcementlearningbasedapproach,10199527}. In \cite{9109717},  \ac{mab}  solutions are explored in ultra-dense mmWave networks, focusing on spatial and temporal learning from empirical mobility trajectories and blockage patterns. While effective in reducing unnecessary HO, these approaches lack real-time context awareness and do not incorporate predictive modeling or physical-layer enhancements. 

We advocate for the adoption of a promising new technology, \ac{riss}, which are programmable surfaces that can selectively reflect and redirect wireless signals and have the capability to improve coverage and reliability, especially in \ac{nlos} scenarios  \cite{basar2019wireless}. In 3D networks involving terrestrial, aerial, and satellite layers, the usage of \ac{riss} has been shown to maintain seamless connectivity \cite{twc2024.3471393}. Cooperative HO strategies that combine \ac{ris} and gNB coordination could further enhance performance, although the cited article \cite{arxiv2407.18183} focuses specifically on reducing signaling overhead caused by RIS reconfiguration during HO management.

\subsection{Contribution}
We aim to address the underexplored intersection of \ac{ris}, mobility prediction, and AI-driven HO logic, offering a scalable solution for dense deployments that aligns with current 3GPP specifications. The main contributions are:

\begin{enumerate}
\item \textbf{Learning-based decision engine:} We develop a decision engine based on \ac{cmab} to select the optimal handover target based on contextual features (e.g. \ac{rsrp}, UE speed, historical performance). \ac{cmab} is preferred over supervised learning techniques due to its low computational cost, adaptability, and robustness in sparse-reward scenarios.
\item \textbf{Predictive modeling:} We implement an \ac{lmmse} predictor to forecast future \ac{rsrp} values, leveraging the temporal correlation of shadowing and path loss. Linear prediction is preferred over neural networks, which tend to be data-wise inefficient in dynamic scenarios 
\item \textbf{\ac{ris}-assisted \ac{ho}:} We propose an optimization procedure to configure \ac{ris} coefficients in coverage-challenged areas. Additionally, we define a method to integrate \ac{ris} into the cell search process using predefined \ac{ris} beam patterns that enable \ac{ssb} detection in harsh propagating conditions.
requiring real-time decisions.
\item \textbf{System integration and evaluation:} We propose a modular architecture that integrates the three subsystems above and evaluates its performance using a comprehensive set of metrics.  
\end{enumerate}


\section{\ac{ltm} \ac{ho} protocol} \ac{ltm} \ac{ho}
 was introduced in 3GPP Release 18 to revolutionize \ac{ho} procedures in 5G-Advanced by enabling URLLC through cell switching in MAC and PHY layers, bypassing traditional RRC-based signaling \cite{gundogan2023ltm}. The procedure avoids resetting entire stacks such as RRC, PDCP, or RLC. It uses preestablished target cell configurations and relies on faster L1/L2 triggering mechanisms. This approach significantly reduces \ac{ho} execution time (under 5 ms), \ac{ho} latency (up to 70\%), signaling overhead (since no RRC commands are expected), and packet loss by reducing reinitialization \cite{10744020}. 
 
 LTM has been shown to excel in many of the following metrics and events that are typically used to evaluate the performance of any HO system:

\begin{itemize}
    \item \textbf{\ac{ho} rate:} The number of \ac{ho}s executions completed per user over a time period. 
  
    \item \textbf{\ac{rlf}}: An \ac{rlf} occurs when \texttt{N310} consecutive out-of-sync indications are received, signaling that channel quality is deteriorating.
    
    \item \textbf{\ac{hof} failures}: Declared when random access to the target cell is not successful.
    
    \item \textbf{\ac{pp} events}: Occur when a UE successfully hands over to a target cell but returns to the previous serving cell within 1000 ms.
    
    \item \textbf{Reliability}: Measured as the ratio of total outage time to total service time. Outages may result from both failed and successful \ac{ho}s.
    
    \item \textbf{Cell preparation events}: It denotes signaling overhead on the radio interface between the UE and the serving cell, as well as the overall network signaling load.
    
    \item \textbf{Resource reservations}: During HO preparation, the target cell reserves resources for the UE. It is quantified as the time between cell preparation and release for a UE.
\end{itemize}

\section{CMAB-based HO decisions} 
\label{RL based HO decisions}
In dynamic 6G environments, traditional HO methods based on signal strength and hysteresis margins fall short due to rapidly changing UE speed and interference. To address this limitation, we introduce a \ac{cmab} approach that provides a lightweight and sample-efficient alternative to full reinforcement learning. The CMAB operates under the assumption that actions (specifically, the selection of the target gNB) do not impact the context vector, which contains observable features: user speed, location,  RSSI, and \ac{sinr}. Optimal actions are selected to maximize long-term rewards related to throughput or HO latency \cite{li2010contextual}.
\subsection{The contextual model}
Let us define the set of actions $\mathcal{A}$, each element $a\in\mathcal{A}$, corresponds to the selection of a specific target gNB. We also define the reward $r\in\mathcal{R}$ as the real value associated to the adopted action. The main goal of applying a \ac{cmab} is to make a sequence of HO decisions (actions $a$) that maximize long-term link quality and reliability based on real-time contextual information, $\mathbf{x}^{(a)}_{t} \in \mathcal{X}  \subseteq \mathbb{R}^d$ that contains HO relevant metrics. 
At each time step $t$, the HO agent observes a context vector $\mathbf{x}^{(a)}_{t}$ describing the environment and selects an action $a_t$ following a strategy that tries to minimize the long term regret, defined as the sum over time of the expected difference between the reward for the best possible action $r^*_t$ and actual reward observed for action $a_t$: $l_t = \sum_{\tau=1}^{t}\mathbb{E}\{r^*_t-r_t^{(a)}\}$.

Each context vector $\mathbf{x}_{t}^{(a)}$ is associated with a candidate $a$ and is designed to include the following components:

\begin{itemize}
    \item Signal quality: The RSRP to the gNB $a$ at time $t$.
    \item Network stability: Moving $p$-samples average \ac{mcs} and SNIR value from cell $a$.
    \item UE mobility: Current UE speed.
    \item HO dynamics: Time since last HO and current serving gNB identifier.
\end{itemize}

In the linear CMAB setting, the expected reward of each $a$ is assumed to be a linear function of the context, enabling efficient algorithms such as LinUCB and Thompson Sampling with provable regret bounds \cite{li2010contextual}. However, real-world rewards may in general be nonlinear functions of the context. One prominent approach to handle nonlinearity is through kernel methods. Algorithms like KernelUCB and GP-UCB leverage this idea using reproducing kernel Hilbert spaces and Gaussian processes, respectively, to model complex reward functions while maintaining theoretical guarantees on regret \cite{valko2013finite}. After some extensive simulations, the rule adopted is the KernelUCB using arc-cosine function \cite{cho2009}. 

As the agent makes a HO decision $a$, it receives a real-valued reward $r_t^{(a)}$ that contains the combined effect of the quality of the HO decision. 

\subsection{Performance-based rewards}
The reward $r_t^{(a)}$ is designed to combine multiple performance indicators observed on the network (see section II):
\begin{align}
r_t^{(a)} =
 r_{\text{thr}} \cdot (1-\mathbb{I}_{\mathrm{RLF}_t})\cdot 
\alpha_{\text{HOF}}^{\mathbb{I}_{\mathrm{HOF}_t}} \cdot \alpha_{\text{HO}}^{\mathbb{I}_{\mathrm{HO}_t}} \cdot  \alpha_{\text{PP}}^{\mathbb{I}_{\mathrm{PP}_t}}
\label{eq:reward_function}
\end{align}
where each term has the following meaning:

\begin{itemize}
    \item $r_{thr}$: Average \ac{mcs} throughput over the last $100\:ms$.    
    \item $\mathbb{I}_{\mathrm{HO}_t}$, $\mathbb{I}_{\mathrm{HOF}_t}$, $\mathbb{I}_{\mathrm{RLF}_t}, \mathbb{I}_{\mathrm{PP}_t}$: Indicator functions that equal 1 if HO, HOF, \ac{rlf} or PP events occurred at time $t$, and 0 otherwise.
    \item $\alpha_i$: Weighting coefficients that balance the relative contribution of each reward component, enabling trade-offs between competing objectives. These are selected experimentally to maximize overall system performance, $\alpha_{HOF} =0.1, \alpha_{HO} = 0.8, \alpha_{PP} = 0.9$.
\end{itemize}



\section{\ac{ltm} based on predicted \ac{rsrp} levels} \label{LTM based on predicted}
To enhance HO reliability in dynamic mobile environments, we propose a predictive strategy that forecasts future \ac{rsrp} values using \ac{lmmse} estimation \cite{9145371}. By leveraging temporal correlations of the pathloss and shadowing in RSRP, the system can pre-reserve resources in target cells offering more stable connectivity, reducing unnecessary signaling and improving HO efficiency.

The processing pipeline consists of the UE measuring raw \ac{rsrp} values from \ac{ssb}s. These raw measurements undergo two layers of filtering —Layer 1 (L1) and Layer 3 (L3)— to mitigate the impact of fast fading and transient fluctuations. The L3-filtered \ac{rsrp} values, which reflect more stable signal trends, serve as the input for the predictor. The \ac{lmmse} prediction horizon $\Delta$ must lie between the decorrelation time of multipath fading (fast variations) and that of shadowing (slow variations). For the \ac{rsrp} values received from every gNB, we compute: 
\begin{equation}
\hat{s}_{t+\Delta} = w_0+\mathbf{w}^T \mathbf{s}_t,
\end{equation}
where $\mathbf{s}_t$ is the vector containing current and past L3-filtered \ac{rsrp} values for a given gNB, and $\{w_0,\mathbf{w}\}$ are the weights that minimize prediction error for that gNB. The covariance matrix and covariance vector of $s_t$ involved are estimated using an exponential window, thus enabling time adaptation. The optimum $\Delta$ can be computed on line following the standard LMMSE solution. 

\section{RIS-assisted HO procedure}
RIS can enhance wireless communication by intelligently reflecting signals toward UE, especially at the cell edge, thereby improving coverage and signal quality \cite{basar2019wireless}. This capability enables more stable connections, facilitates better HO decisions, and reduces RLF and PP effects. 
\subsection{Channel model}
For \ac{ris}-assisted links, the cascaded MISO channel gain for the compound link gNB$\rightarrow$\ac{ris}$\rightarrow$UE is expressed as: 
\begin{equation}
    \label{eq:compund_channel}
    \mathbf{H}_c=\mathbf{h}_{ru}^T\mathbf{\Theta}\Hbr,
\end{equation}
where \(\Hbr\in\mathbb{C}^{N_r\times M}\) contains the channel gains between the $M$ antennas at the gNB and the $N_r$ RIS elements, and \(\mathbf{h}_{ru}\in\mathbb{C}^{N_r\times 1}\) contains the channel gains between the \ac{ris} and a single antenna UE. Moreover, $\mathbf{\Theta} \in \mathbb{C}^{N_r\times N_r}$ is the \ac{ris} scatter matrix. For the diagonal passive \ac{ris} case, the unit-modulus reflection coefficients are stacked in $\mathbf{\Theta}=diag(e^{j\pmb{\phi}})\in \mathbb{C}^{N_r\times N_r}$, where $\pmb{\phi}=\left [\phi_1 \: \phi_2 \: \cdots \: \phi_{N_r}  \right ]$ are phase shifts applied to each \ac{ris} element independently reflecting the incident signal.

\subsection{Optimization of \ac{ris} scatter matrix}
When the UE establishes a connection with the gNB, conventional approaches assume that full channel state information (CSI) can be obtained. In our \ac{ris}-assisted scenario case, we want the \ac{ris} to reflect the \ac{ssb} signal omnidirectionally transmitted from a single antenna by a gNB operating without explicit knowledge of the \ac{ris}$\rightarrow$UE channel state. To address this issue, predefined \ac{ris} beam patterns are pre-computed to reflect \ac{ssb}s across an $N \times N$ area, enabling UEs to detect signals via indirect paths even when direct links are blocked. This approach supports initial access and synchronization in 5G \ac{nr} systems, where \ac{ssb}s are periodically transmitted using single antenna panels.

\begin{figure}[h]
    \centering
    \begin{subfigure}{0.23\textwidth}
        \includegraphics[width=\linewidth]{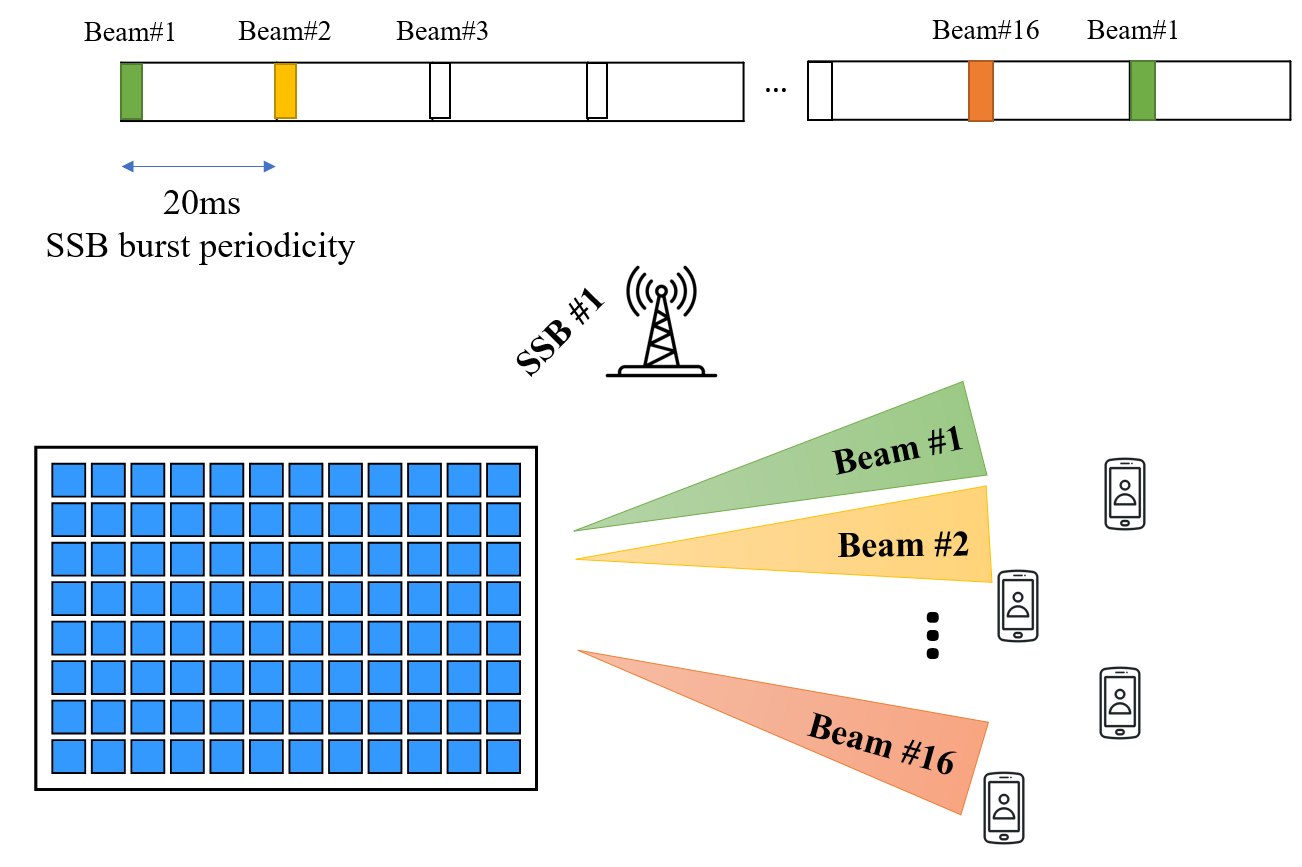}
        \caption{Case 1: Sequential reflection}
        \label{fig:1BeamSSBburst}
    \end{subfigure}    
    \begin{subfigure}{0.23\textwidth}
        \includegraphics[width=\linewidth]{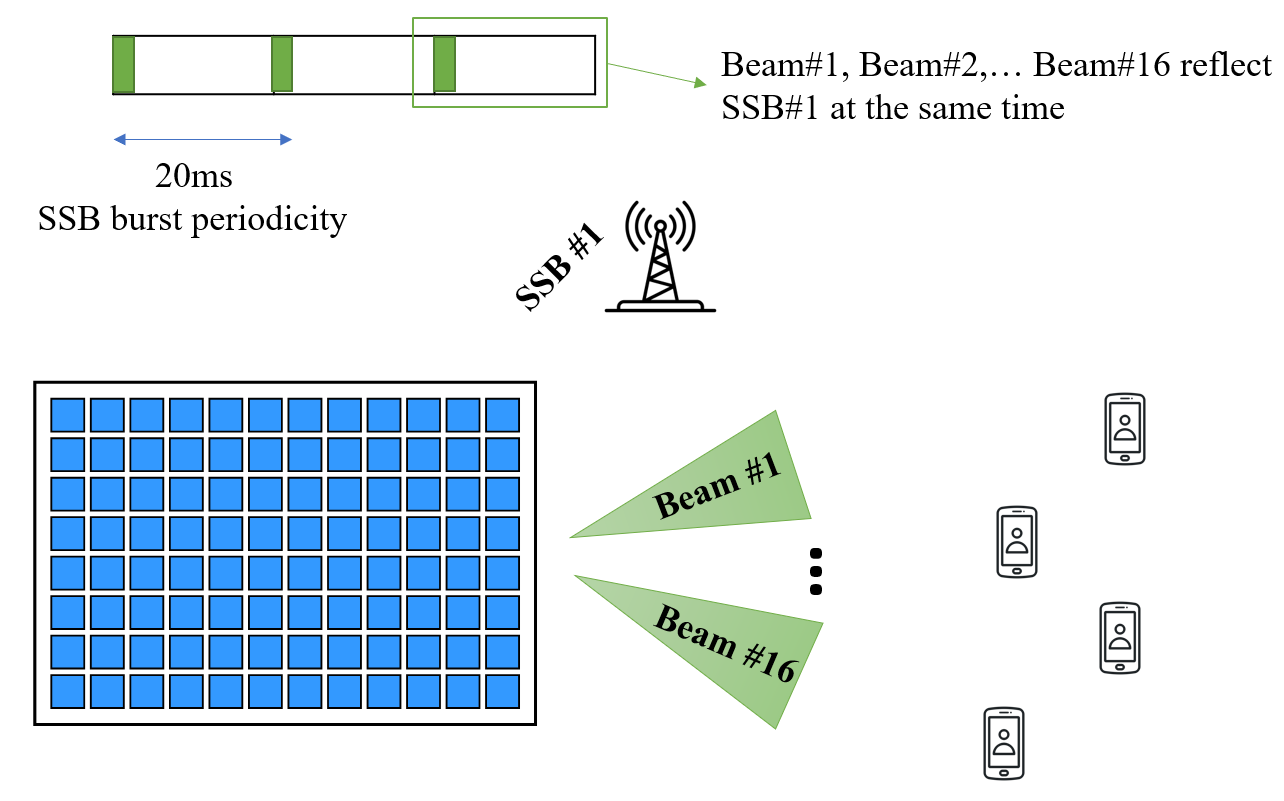}
        \caption{Case 2: Simultaneous reflection}
        \label{fig:16BeamsSSBburst}
    \end{subfigure}
        
    \caption{Two proposed \ac{ris}-assisted transmission of SSB frames}
\label{fig:ComparisonRISOptimization}
\end{figure}

Two different approaches for the use of \ac{ris} beams are considered here. The first consists of sequentially applying different predefined \ac{ris} beam patterns per \ac{ssb} transmission frame. Over multiple time multiplexed frames, the \ac{ris} sweeps through $T$ directions sequentially (see Fig. \ref{fig:1BeamSSBburst}).

The second approach consists of simultaneously applying several \ac{ris} beam patterns to each \ac{ssb} time frame. A fraction of the \ac{ris} elements synthesizes each beam and covers a different portion of the area, providing faster \ac{ssb} identification but with lower received power (Fig. \ref{fig:16BeamsSSBburst}).

In both approaches, to optimize \ac{ris} reflection performance over a target $N \times N$ area, we propose a worst-case zone-driven optimization approach. The method assumes a set of hypothetical user locations uniformly spread across the area. It determines the \ac{ris}$\rightarrow$UE channels based on geometric layout, \ac{los} propagation, and electromagnetic reflection models.

We denote the received signal power at location $b$ as $|h_{\text{eff}}^{(b)}|^2$, which captures the effective gain of the gNB$\rightarrow$\ac{ris}$\rightarrow$UE link. Since we do not have UE-specific CSI, the estimated received power is computed using:
\begin{equation}
\label{eq:power_r}
P_r^{(b)}=|h_{\text{eff}}^{(b)}|^2 = |\mathbf{h}_{\text{ru}}^{(b)T} \boldsymbol{\Theta} \mathbf{h}_{\text{br}}|^2,
\end{equation}
where vector $\mathbf{h}_{br}$ contains the channel gains between the antenna at the gNB transmitting \ac{ssb} omnidirectionally and the \ac{ris} elements. The key idea is to maximize the minimum gain over the spatial grid, effectively guaranteeing robust worst-case performance. The optimization goal becomes:
\begin{equation}
\label{eq:heff}
\max_{\pmb{\phi}} \min_{b \in \mathcal{B}} P_r^{(b)},
\end{equation}
where $\mathcal{B}$ is the set of all discrete locations covering the $N \times N$ grid area. This \emph{max-min fairness} criterion ensures that the most shadowed zone in the area is enhanced as much as possible, under the principle that coverage over the entire area is more critical than optimizing for individual high-gain users. We will adopt a gradient approach to determine the best \ac{ris} coefficients. The maximization of $P_r^{(b)}$ reads:
\begin{equation}
\begin{aligned}
\max_{\pmb{\phi}} \: & \left| \mathbf{h}_{\text{ru}}^{(b)T}\boldsymbol{\Theta} \mathbf{h}_{\text{br}} \right|^2 \\
&\text{s.t.} \quad \phi_i \in [0, 2\pi), \ \forall i = 1, \dots, N_r
\end{aligned}
\end{equation}
The gradient of $P_r$ with respect to \ac{ris} phases $\pmb{\phi}$ is obtained from complex matrix differentiation:
\begin{equation}
    \label{eq:gradient}
    dP_r = \text{Im} \left[h_\text{eff} \mathbf{H}_{12}\textup{diag}(e^{j\pmb{\phi}})\right ]d\pmb{\phi}=\mathcal{D}_{\pmb{\phi}}d\pmb{\phi},
\end{equation}
where $\mathbf{H}_{12}=[vec(\mathbf{h}_{ru,1}\mathbf{h}_{br,1}^T) \cdots vec(\mathbf{h}_{ru,N_r}\mathbf{h}_{br,N_r}^T)]$ and Im is the imaginary part. From \eqref{eq:gradient}, the gradient vector $\mathcal{D}_{\pmb{\phi}}$ can be read out. Algorithm \ref{alg:RIS_Optimization_Adam} iteratively optimizes the \ac{ris} coefficients via the Adam gradient method. At each iteration the gradient of the power at the position $b$ with the minimum $P_r^{(b)}$ is selected.

\begin{algorithm}[h!]
\caption{Iterative optimization of \ac{ris} coefficients}
\label{alg:RIS_Optimization_Adam}
\begin{algorithmic}
\State $\textbf{input: } \mathbf{h}_{br}, \mathbf{H}_{ru}, \mathcal{B}, \varepsilon = 1e{-}4$ 
\State $\textbf{output: } \pmb{\varphi}$ 
\State compute $\:\mathbf{H}_{12}$
\State Initialize $t \gets 0,\:J_0\gets0$,\:$\pmb{\phi}_0\gets \text{random},\:\pmb{\varphi}=e^{j\pmb{\phi}_0}$
\Repeat
\State compute $\:\{P^{(b)}\}_{b\in \mathcal{B}}$ \hfill\Comment{Power received at locations in $\mathcal{B}$ \eqref{eq:power_r}}
\State $b_{m} = \argmin_{b \in \mathcal{B}} P^{(b)}$ \hfill \Comment{Select location with minimum $P_r$}
\State $J_t = P_r^{(b_m)}$ \hfill \Comment{Received power metric}
\State $t \gets t+1$
\State $\mathbf{g}_t=D_{\pmb{\phi}}J_{t-1}$ \hfill \Comment{Compute gradient using \eqref{eq:gradient}}
\State $\pmb{\phi}_t = \text{AdamGradientUpdate}(\mathbf{g}_t,\pmb{\phi}_{t-1})$
\State $\pmb{\varphi}=e^{j\pmb{\phi}_t}$
\Until {$|\frac{J_t-J_{t-1}}{J_t}|<\varepsilon$}
\end{algorithmic}
\end{algorithm}

\section{Results} 
\label{RL based HO decisions}
We compare the proposed \ac{ris}-assisted, CMAB-optimized HO strategy with the traditional \ac{rss}-based handover commonly used in LTE/5G systems.
\subsection{Simulation setup}
The simulation parameters are divided into three main categories based on the system modules, each configured to capture realistic 5G urban deployment scenarios.

\paragraph{Channel Model and Scenario}
Three channels are considered: the direct channel gNB$\rightarrow$UE ($\mathbf{h}_{bu}$), the channel gNB$\rightarrow$\ac{ris} ($\mathbf{h}_{br}$), and the channel \ac{ris}$\rightarrow$UE ($\mathbf{h}_{ru}$). Models for $\mathbf{h}_{ru}$ and \(\Hbr\) have been described in \cite{10437329}. The direct channel model follows the 3GPP TR 38.901 specifications \cite{3gpp_38901} for urban macro environments. In regions covered by the RIS, an additional attenuation factor is applied to the gNB$\rightarrow$UE link to account for signal degradation of 20dB due obstruction.

In contrast, the gNB$\rightarrow$\ac{ris} and \ac{ris}$\rightarrow$UE channels are modeled with large-scale path loss (from the 3GPP TR 38.901) under the assumption that both links operate under LOS. The gNB$\rightarrow$RIS link is static and unobstructed, while the \ac{ris}$\rightarrow$UE link is assumed to be dominated by the deterministic geometry of the \ac{ris} deployment and the beam configuration.

A total of 7 cells are placed on a hexagonal grid, while two \ac{ris} panels per cell are deployed on cell edges to enhance RSRP in low coverage zones. Each cell is sectored at $120^\text{o}$ with a frequency reuse of 1/3. Other parameters are listed in Table \ref{tab:sim_param:channel_network}.
\begin{table}[t]
\centering
\caption{Channel and Network Topology Parameters}
\label{tab:sim_param:channel_network}
\begin{tabular}{|l|l|}
\hline
\textbf{Parameter}                       & \textbf{Value / Description}           \\ \hline
Path loss model                          & LOS 3GPP UMa in TR 38.901               \\
Shadowing standard deviation             & 4 dB                                   \\
Small-scale fading                       & Rayleigh-Jakes model \\
Carrier frequency                        & 10 GHz                                 \\
Bandwidth                                & 200 MHz                                \\
UE speed                                 & 10–18 km/h (running pedestrian)        \\ \hline
Number of BSs                            & 7 (hexagonal cell)                     \\
Inter BS Spacing                         & 200 m                                  \\
Sectors per BS                           & 3 (each 120$^\circ$)                   \\
BS antenna gain                             & 3 dBi                                  \\
$\theta_{\text{3dB}}, \phi_{\text{3dB}}$ & 65$^\circ$                             \\
Maximum attenuation ($A_m$)              & 30 dB                                  \\
Antenna element spacing                  & $\lambda/2$                            \\
Transmit power                           & 25 dBm                                 \\
Noise power density                      & –174 dBm/Hz                            \\
Frequency reuse factor                   & 3                                      \\ \hline
\end{tabular}
\end{table}

\paragraph{\ac{ris} Setup} 
Each \ac{ris} features a total of $N_r=12800$ elements. The strategy considered is the single-beam \ac{ssb} reflection over a \(40\text{m} \times 40\text{m}\) area that will be denoted as $\mathcal{C}$. This area is partitioned into 16 sub-areas, each denoted by \(\mathcal{B}_i\), \(i = 1, 2, \dots, 16\), corresponding to square regions of \(10\text{m} \times 10\text{m}\). Furthermore, each sub-area \(\mathcal{B}_i\) is subdivided into \(100\) smaller squared regions \(b_j\), \(j = 1, 2, \dots, 100\) of \(1\text{m} \times 1\text{m}\). To optimize the \ac{ris} configuration over $\mathcal{C}$, Algorithm \ref{alg:RIS_Optimization_Adam} is applied independently to each sub-zone $\mathcal{B}_i$, so a codebook of 16 RIS coefficient sets is used to reflect the \ac{ssb} according to the schemes in Fig. \ref{fig:ComparisonRISOptimization}.  The two \ac{ris}s placed at each cell are in diagonal opposites. RIS$_1$ is located at coordinates $  [-40\text{m},-40\text{m}]$ while RIS$_2$ is located opposite to RIS$_1$, at $[+40 \text{m},+40\text{m}]$ relative to the cell center.

\begin{figure}[h!]
    \centering
    \begin{subfigure}[b]{0.23\textwidth}
   \includegraphics[width=\linewidth]{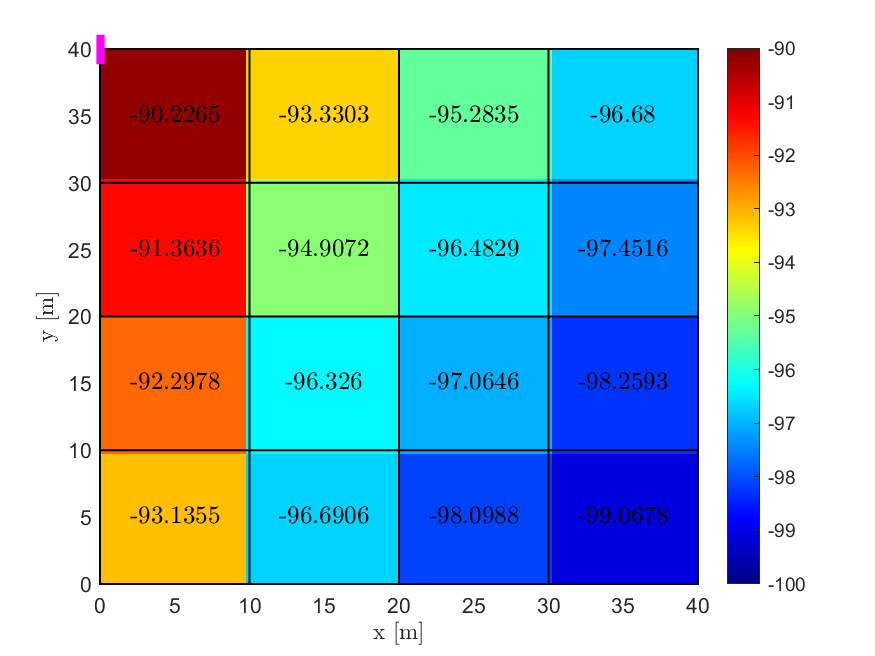}
    \caption{Sequential beam sweeping}
    \label{fig:1BeamSSBburst_Zone}
    \end{subfigure}
    \hfil
    \begin{subfigure}[b]{0.23\textwidth}
    \includegraphics[width=\linewidth]{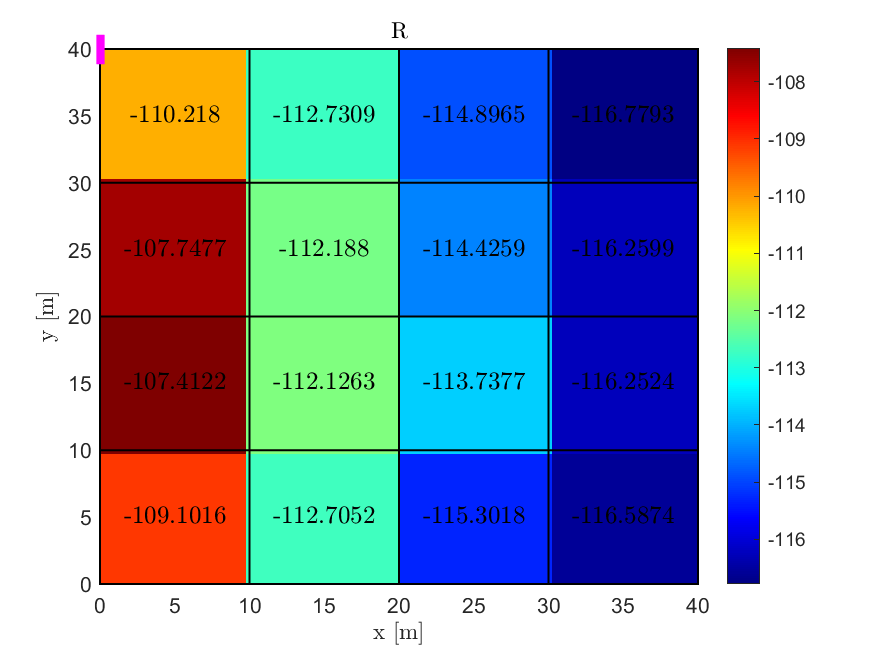}
    \caption{Simultaneous beam sweeping}
    \label{fig:16BeamsSSBburst_Zone}
    \end{subfigure}
    
    \caption{Average received power per $10\text{m}\times 10\text{m}$ square over a total $40\text{m} \times40\text{m}$ area assisted by a \ac{ris}, using the two proposed strategies (\ac{ris} is located at the top-left corner of the area)}
    \label{fig:RIS_Strategies_Pr}
\end{figure}

Fig. \ref{fig:RIS_Strategies_Pr} shows the \ac{ssb} received signal power in the area covered by the \ac{ris}, a tiny 9.2\% of the total cell area. Fig. \ref{fig:1BeamSSBburst_Zone} uses sequential single-beam sweeping, where the $12.800$ \ac{ris} elements steer power on one sub-zone at a time. In contrast, in Fig. \ref{fig:16BeamsSSBburst_Zone}, simultaneous beam sweeping is employed: each of the 16 sub-zones is focused by 800 \ac{ris} elements. Results show that the single-beam sequential strategy from Fig. \ref{fig:1BeamSSBburst_Zone} achieves significantly higher received signal power, with gains exceeding those of the simultaneous case by 10–15 dB. On the contrary, sequential reflection entails a tradeoff between signal stability and HO latency that is left for further study.

\paragraph{Handover Parameters} 
The \ac{ho} procedure includes \ac{ltm} signaling procedures, prediction-based triggers using \ac{lmmse}-predicted RSRP, and penalization of ping-pong and failed handovers. UEs perform periodic SSB measurements, and events such as \ac{ho}, \ac{hof}, and \ac{rlf} are tracked. Statistical evaluation includes averaging over 1.000 independent simulated UE trajectories over 300 s each. The rest of the parameters associated with \ac{ho} are displayed in Table \ref{tab:HO_params}.

\begin{table}[H]
\centering
\caption{Handover Simulation Parameters}
\begin{tabular}{|l|l|}
\hline
\textbf{Parameter}             & \textbf{Value / Description} \\ \hline
Prediction Delay               & 20 s                         \\
HO offset threshold            & 3 dB                         \\
Target cell pre-selection      & Predicted top-$4$ neighbors  \\
Receiver sensitivity           & –95 dBm                      \\
Time resolution (simulation)   & 10 ms                        \\
Simulation duration            & 300 s                        \\
Number of UEs                  & 1000 mobile UEs               \\
L1/L3 Measurement Report Delay & 10 ms                        \\
HO Decision Computation        & 10 ms                        \\ \hline
\end{tabular}
\label{tab:HO_params}
\end{table}



\begin{figure}[h!]
    \centering
    \begin{subfigure}[b]{0.2\textwidth}
        \includegraphics[width=\textwidth]{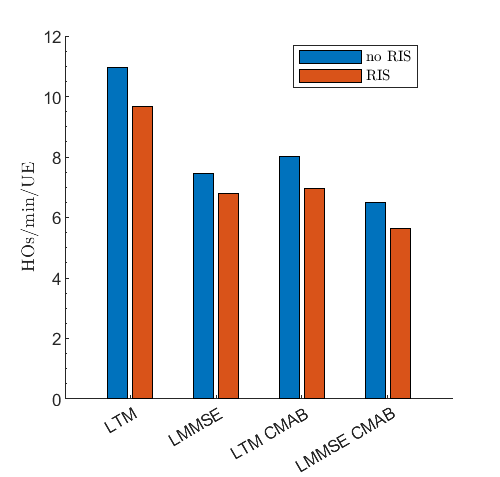}
        \caption{HO/min}
        \label{fig:ALL:HO}
    \end{subfigure}
    \hfil
    \begin{subfigure}[b]{0.2\textwidth}
        \includegraphics[width=\textwidth]{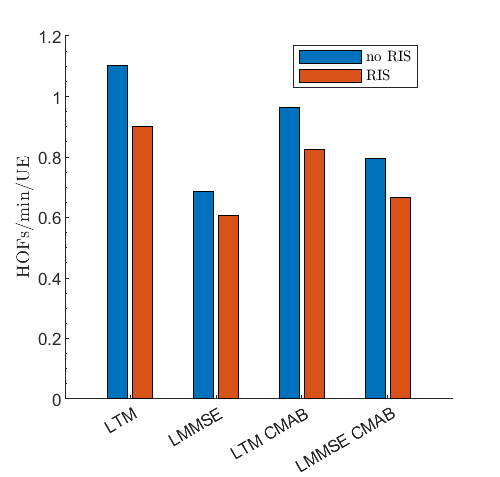}
        \caption{HOF/min}
        \label{fig:ALL:HOF}
    \end{subfigure}
    
    \vspace{0.5cm}
    
    \begin{subfigure}[b]{0.2\textwidth}
        \includegraphics[width=\textwidth]{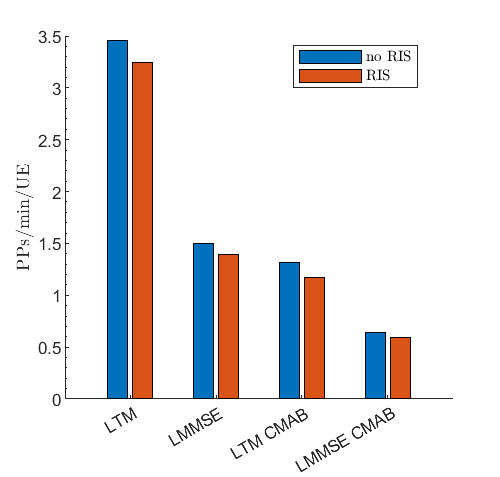}
        \caption{PP/min}
        \label{fig:ALL:PP}
    \end{subfigure}
    \hfil
    \begin{subfigure}[b]{0.2\textwidth}
        \includegraphics[width=\textwidth]{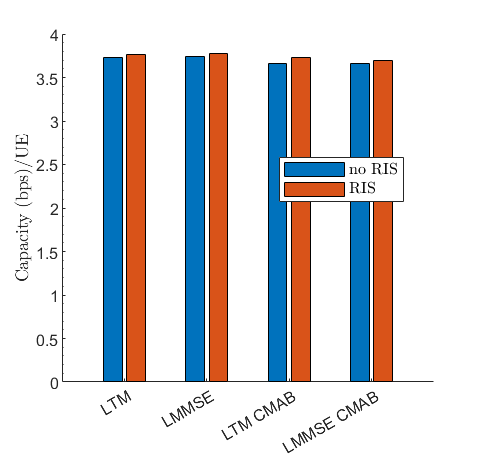}
        \caption{Capacity (bps)}
        \label{fig:ALL:Cap}
    \end{subfigure}

    \begin{subfigure}[b]{0.2\textwidth}
        \includegraphics[width=\textwidth]{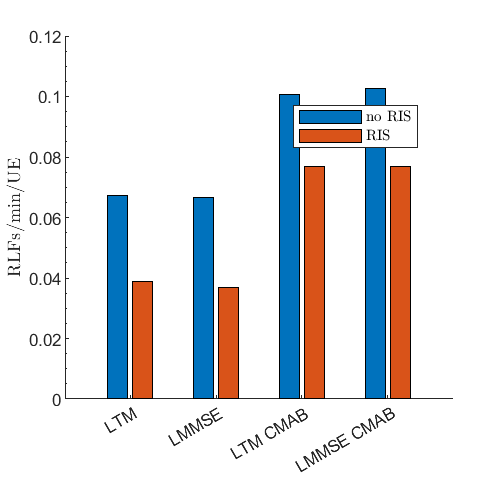}
        \caption{RLF/min}
        \label{fig:ALL:RLF}
    \end{subfigure}
    \hfil
    \begin{subfigure}[b]{0.2\textwidth}
        \includegraphics[width=\textwidth]{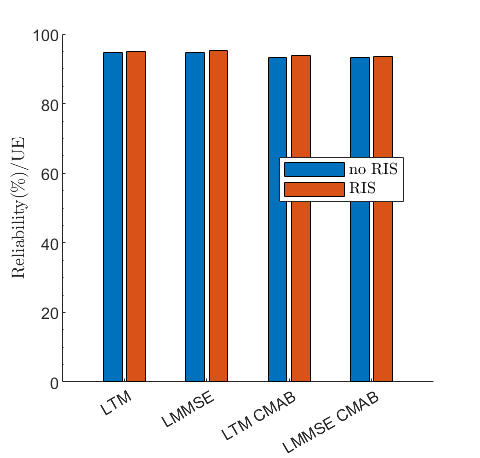}
        \caption{Reliability (\%)}
        \label{fig:ALL:Reliability}
    \end{subfigure}
    
    \vspace{0.5cm}
    \begin{subfigure}[b]{0.2\textwidth}
        \includegraphics[width=\textwidth]{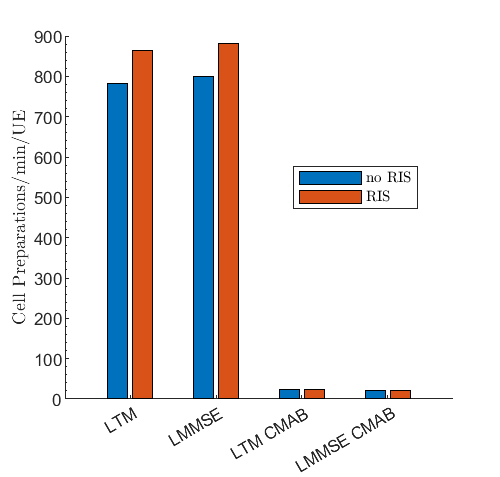}
        \caption{Cell preparations/min}
        \label{fig:ALL:CellPreparations}
    \end{subfigure}
    \hfil
    \begin{subfigure}[b]{0.2\textwidth}
        \includegraphics[width=\textwidth]{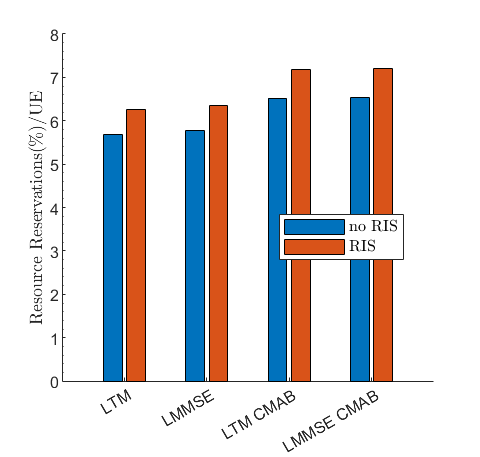}
        \caption{Resource reservations (\%)}
        \label{fig:ALL:RR}
    \end{subfigure}
    
    \caption{Comparison of HO metrics. Each plot displays one HO-related KPI from section II, for the eight proposed HO configurations.}
    \label{fig:ALL_1}
\end{figure}

\subsection{Comparative performance}
To evaluate the effectiveness of each proposed techniques in HO performance, we analyze the eight configurations:

\begin{enumerate}
    \item LTM
    \item LTM + RIS
    \item LTM + LMMSE
    \item LTM + LMMSE + RIS
    \item LTM + CMAB
    \item LTM + CMAB + RIS
    \item LTM + CMAB + LMMSE
    \item LTM + CMAB + LMMSE + RIS
\end{enumerate}
under the same environmental and mobility conditions to allow fair comparison. The ultimate goal is to determine which combination yields the best performance across the multiple KPIs described in Section II. Conclusions from the results in Fig. \ref{fig:ALL_1} are compiled below:

\vspace{0.5em}
\noindent- \textbf{\ac{cmab} Decisions:}  
\ac{cmab} reduces the frequency of \ac{ho}s, \ac{pp} events, \ac{hof}s (at a lesser extent), and significantly cuts down on cell preparation overhead by leveraging contextual information. However, it leads to a slighly higher resource reservations and increased \ac{rlf}s due to delayed exploratory decisions. When combined with predictive (\ac{lmmse}) or physical-layer (\ac{ris}) enhancements, reliability remains consistently competitive.

\vspace{0.5em}
\noindent- \textbf{RSRP Prediction:}  
\ac{lmmse} forecast of RSRP improves handover reliability and reduces \ac{rlf}s by anticipating channel degradation. Its benefits are particularly significant in the absence of \ac{cmab}. When \ac{cmab} is applied, \ac{lmmse} contribution becomes marginal, as \ac{cmab} already incorporates temporal and contextual trends in the decision-making process.

\vspace{0.5em}
\noindent- \textbf{\ac{ris} Integration:}  
\ac{ris} strengthens link quality and reduces \ac{rlf}s by providing alternative reflective paths. This results in improved \ac{ho} success as it also reduces PP rates and number of HOF. However, it can slightly increase the amount of cell preparations and resource reservations. Still, \ac{ris} consistently enhances signal reliability and reduces \ac{rlf}.

\subsection{Best configurations based on LTM HO}
\label{sec:best_strategies}
The configuration combining \textbf{\ac{ltm} + \ac{lmmse} + \ac{ris} + \ac{cmab}} offers the best overall performance by balancing high reliability, reduced \ac{hof}s and \ac{rlf}s, and optimized \ac{ho}s decisions with minimal control-plane overhead. It also achieves the lowest \ac{pp} rate, benefiting from predictive awareness and \ac{ris}-enhanced link support, while accepting slightly higher resource reservations as a trade-off. For reliability-focused scenarios, \textbf{\ac{ltm} + \ac{lmmse}} and \textbf{\ac{ltm} + \ac{lmmse} + \ac{ris}} stand out, while \textit{\ac{cmab}-based} approaches —especially those integrating \ac{ris}— excel in minimizing HO/min and ping-pongs. Additionally, all \textit{\ac{cmab}-integrated} setups effectively reduce cell preparation overhead by targeting relevant cells.

\section{Conclusions}
This study presents a comprehensive evaluation of advanced HO strategies in 5G systems by integrating multiple technological innovations in a baseline \ac{ltm} procedure. \ac{lmmse} prediction shows enhancement in decision timing, allowing \ac{ho}s to be initiated before critical link degradation, \ac{ris} integration boosts signal propagation and reliability and \ac{cmab} provides better HO decision-making. This combination results in a highly effective strategy that reduces unnecessary \ac{ho}s, link failures (both \ac{hof}s and \ac{rlf}s), ping-pongs, and cell preparations.
\bibliographystyle{IEEEtran} 
\bibliography{mybiblio_clean}

\end{document}